\documentclass[journal, onecolumn, draftcls,12pt]{IEEEtran}
\usepackage{amssymb,amsmath,epsfig,graphicx,theorem}
\usepackage{amsfonts}
\usepackage{bm}
\newtheorem{Theo}{Theorem}

\usepackage{epsfig,rotating,setspace,latexsym,amsmath,epsf,amssymb,bm,color}
\usepackage{cite}

\DeclareMathOperator{\sgn}{sgn}

\IEEEoverridecommandlockouts

\title{An Upper Bound on the Sum Capacity of the Downlink Multicell Processing with Finite Backhaul Capacity}

\author{Tianyu Yang, Nan Liu, Wei~Kang, and Shlomo Shamai (Shitz)%
\thanks{T. Yang and W. Kang are with the Information Security Research Center,
Southeast University, Nanjing, China (email: \{tianyu,wkang\}@seu.edu.cn). N. Liu is with the National Mobile Communications Research Laboratory,
Southeast University, Nanjing, China (email: nanliu@seu.edu.cn). S. Shamai (Shitz) is with the Department of Electrical Engineering, Technion Israel Institute of Technology, Haifa 32000, Israel (e-mail: sshlomo@ee. technion.ac.il).}%
}

\begin{document}

\maketitle
\begin{abstract}
In this paper, we study upper bounds on the sum capacity of the downlink multicell processing model with finite backhaul capacity for the simple case of 2 base stations and 2 mobile users. It is modeled as a two-user multiple access  diamond channel. It consists of a first hop from the central processor to the base stations via orthogonal links of finite capacity, and the second hop from the base stations to the mobile users via a Gaussian interference channel. The converse is derived using the converse tools of the multiple access diamond channel and that of the Gaussian MIMO broadcast channel. Through numerical results, it is shown that our upper bound improves upon the existing upper bound greatly in the medium backhaul capacity range, and as a result, the gap between the upper bounds and the sum rate of the time-sharing of the known achievable schemes is significantly reduced.
\end{abstract}

\section{introduction}
 The multi-cell processing system, as reviewed in \cite{O.Simeone:2012}, has been used to increase the throughput and to cope with the inter-cell interference. The downlink multi-cell processing system, when first considered, consists of different base stations linked to the central processor via  backhaul links of \emph{unlimited} capacity, and therefore, the amount of cooperation among the different base stations is unbounded. This network can be modeled by a MIMO broadcast channel and the sum-rate characterization was found in \cite{O.Somekh:2007}. Later on, due to the impracticality of unlimited capacity backhaul links, \cite{O.Simeone:2009,S-HPark:2013,S-NHong:2013,N.Liu:2014,Y.Xiao:2015} studied the problem of finding the capacity region of the downlink multicell processing system when the capacities of the backhaul links are \emph{finite}, and proposed various achievable schemes to efficiently utilize the finite capacity backhaul links.
%  However, in the most practical scenarios, the capacity of backhaul link is limited. In order to increase the whole throughput of this system, the limited backhaul must be fully utilized. Thus, the question of how to transmit correlated codes to different base stations to make best use of the limited backhaul link arises.
%
%    Several schemes of sending correlated codes with finite-capacity backhaul links are studied in [3-6]. 
More specifically, in \cite{O.Simeone:2009}, a compressed dirty-paper coding scheme is proposed, where the base stations are treated as the antennas of the central processor and the dirty-paper coding codewords for each antenna are compressed and transmitted on the backhaul links. The scheme is improved in \cite{S-HPark:2013} by allowing the quantization noise of the base stations be correlated. The scheme of reverse compute-and-forward was proposed in \cite{S-NHong:2013} where linear precoding is performed at the central processor and the backhaul links are used to transmit linear combinations of the messages over a finite field. Such linear precoding transforms the channel seen at each mobile user into a point-to-point channel where integer-valued interference is eliminated by precoding and the remaining noninteger residual interference is treated as noise. By regarding the network model as a multi-user diamond channel,  an  achievability scheme is proposed in \cite{N.Liu:2014, Y.Xiao:2015} by combining
%, which is studied in \cite{W.Kang:2015}. 
Marton's achievability for the broadcast channel \cite{Marton:1979} and the achievability of sending correlated codewords over a multiple access diamond channel \cite{Ahlswede:1983, Traskov:2007}.
% are combined and the achievability scheme is proved better than other schemes in the most cases.

 The outer bound on the capacity region for this network is unknown except for the simple cut-set bound \cite{Cover:book}, which is the minimum of the capacity between the first  hop from the central processor to the base stations and that of the second hop from the base stations to the mobile users. When the capacity of the backhaul links are relatively large, the performance of the scheme of compressed dirty-paper coding approaches that of the simple cut-set bound.  On the other hand, when the capacity of the backhaul links are relatively small, the scheme of reverse compute-and-forward reaches the simple cut-set bound \cite{N.Liu:2014}. In the medium capacity region, 
% even though the scheme proposed by \cite{N.Liu:2014} improves upon the schemes of compressed dirty-paper coding and reverse compute-and-forward, 
 there is still a relatively large gap between the simple cut-set upper bound and the performance of the time-sharing of the known achievable schemes. So it is unknown how well the proposed achievable schemes are and whether further efforts are needed in proposing better achievable schemes than existing ones for the downlink multicell processing system. 
 
 In this paper, we derive a novel upper bound on the sum capacity of the downlink multicell processing network consisting of two base stations and two users. Similar to \cite{N.Liu:2014}, we regard the network as a 2-user multiple access diamond channel. 
% %
% As a result, the converse is derived using the converse tools of the multiple access diamond channel used in \cite{W.Kang:2015, G.Kramer:2014} and that of the Gaussian MIMO broadcast channel used in \cite{C.Nair:2014}. 
 We first provide a cut-set upper bound using more cuts than the known simple cut-set bound of the minimum between the capacities of the first and the second hop.  Next, single-letterization methods for the Gaussian multiple access diamond channel \cite{W.Kang:2015, G.Kramer:2014, G.Kramer:2015, G.Kramer:2016}  is applied to our problem. Finally, we obtain a novel upper bound on the sum capacity utilizing the converse tools of the Gaussian MIMO broadcast channel in \cite{C.Nair:2014}. The derived upper bound is expressed in terms of the sum capacity of the Gaussian MIMO broadcast channel given input covariance constraint, which has been found in \cite{H.Weingarten:2006, T.Liu:2007, C.Nair:2014,Viswanath:2003, Vishwanath:2003, Yu:BC2004}, and thus, is easy to evaluate numerically. 
 
Comparing numerically the proposed upper bound, the simple cut-set upper bound and the sum rate of various achievable schemes for the multicell processing system in terms of the sum-rate, we see that our upper bound improves upon the existing simple cut-set upper bound greatly in the medium backhaul capacity range, and as a result, the gap between the upper bounds and the sum rate of the time-sharing of the known achievable schemes is significantly reduced.

%    The remainder of this paper is organized as follows. In section II, we provide the system model and give some efficient definitions. In section III, we deprive a new upper bound for the capacity region using the method in \cite{W.Kang:2015}. The correlated computation results is in section IV, and the conclusion is in section V.

\section{system model}
\begin{figure}[t!]
\centering
\includegraphics[width=4in]{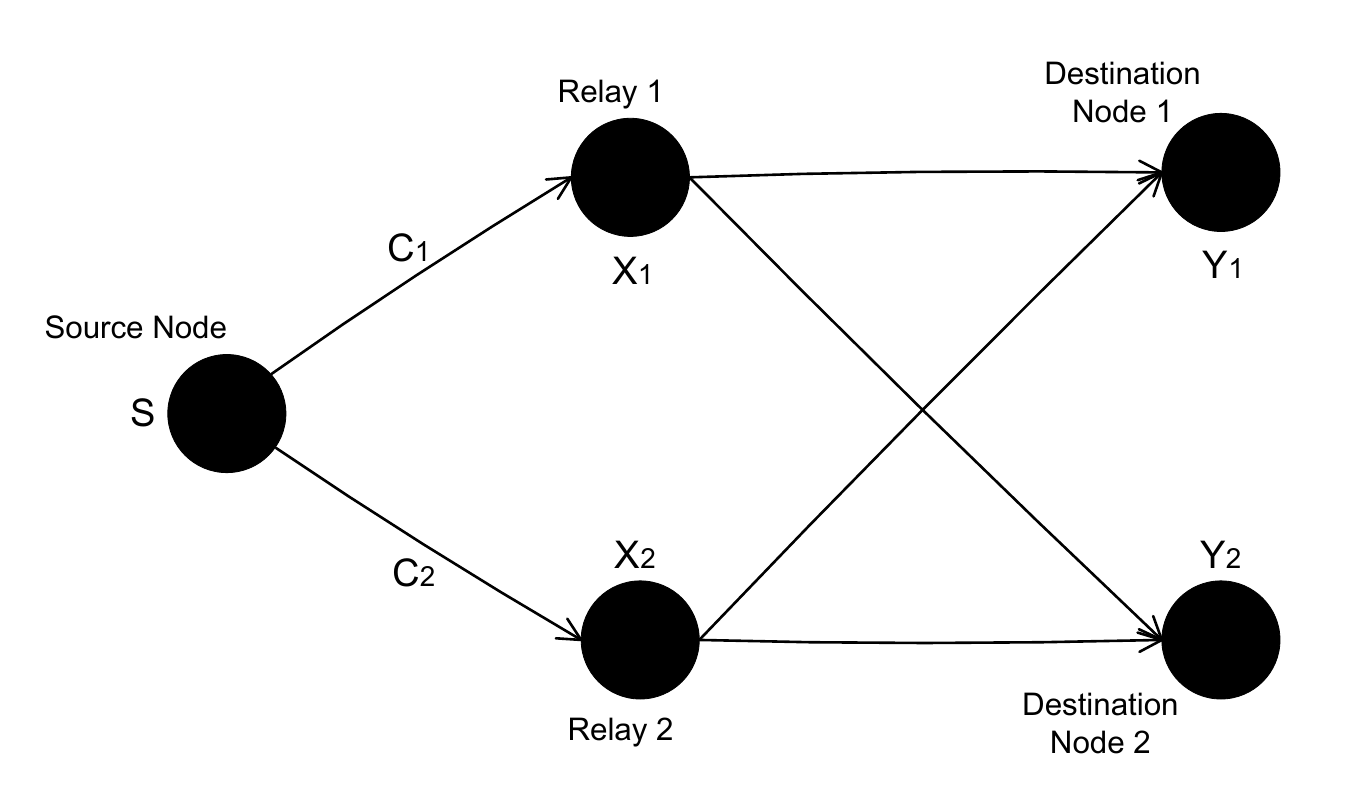}
\caption{The 2-user multiple access diamond channel.} \label{MAC_diamond}
\end{figure}

In this paper, we consider the downlink multicell processing system with two base stations and two users. This network model can be seen as the 2-user multiple access diamond channel \cite{N.Liu:2014}, see Fig. \ref{MAC_diamond}.  The source node (central processor) can transmit to Relays (base stations) 1 and 2  via backhaul links of capacities $C_1$ and $C_2$, respectively. The channel between the two relay nodes and the two destination nodes (mobile users) is characterized by $p(y_1,y_2|x_1,x_2)$, with input alphabets $(\mathcal{X}_1,\mathcal{X}_2)$ and output alphabets $(\mathcal{Y}_1,\mathcal{Y}_2)$. 
%In this paper, we call this channel \emph{2-destinations multiple access diamond chanel}. 
Let $W_1$ and $W_2$ be two independent
messages that the source node would like to transmit to Destinations 1 and 2, respectively. Assume that $W_k$ is uniformly distributed on
$\{1,2,\cdots,M_k\}$, $k=1,2$. 

An $(M_1,M_2,n,\epsilon_n)$ code consists of an encoding function at the source node:
\begin{align}
f^n:\{1,2,\cdots,M_1\}\times\{1,2,\cdots,M_2\} \rightarrow \{1,2,\cdots,2^{nC_1}\}\times\{1,2,\cdots,2^{nC_2}\}, \nonumber
\end{align}
two encoding functions at the relay nodes:
\begin{align}
f_k^n:\{1,2,\cdots,2^{nC_k}\} \rightarrow \mathcal{X}_k, \quad k=1,2, \nonumber
\end{align}
and two decoding functions at the destination nodes:
\begin{align}
g_k^n:\mathcal{Y}_k \rightarrow \{1,2,\cdots,M_k\},\quad k=1,2. \nonumber
\end{align}
The average probability of error is defined as 
\begin{align}
\epsilon_n=\sum_{w_1=1}^{M_1} \sum_{w_2=1}^{M_2}\frac{1}{M_1M_2}Pr[g_1^n(Y_1^n)\neq w_1 \text{ or }g_1^n(Y_2^n)\neq w_2|W_1=w_1,W_2=w_2]. \nonumber
\end{align}
Rate pair $(R_1,R_2)$ is said to be achievable if there exists a sequence of $(2^{nR_1},2^{nR_2},n,\epsilon_n)$ code such that $\epsilon_n \rightarrow 0$ as $n\rightarrow \infty$. The capacity of the 2-user multiple access diamond channel is the closure of the set of all achievable rates pairs.

In this paper, we study the Gaussian case, where $\mathcal{X}_1=\mathcal{X}_2=\mathcal{Y}_1=\mathcal{Y}_2=\mathbb{R}$, and the channel between the two relays and each destination node is a Gaussian multiple access channel, i.e., the received signals at the destination nodes are
\begin{align}
Y_1=X_1+aX_2+U_1, \label{GC01} \\
Y_2=bX_1+X_2+U_2, \label{GC02}
\end{align}
where $X_1$ and $X_2$ are the input signals from Relays 1 and 2, respectively, $U_1$, $U_2$ are two independent zero-mean unit-variance Gaussian random variables that are independent to $(X_1,X_2)$, and $a,b \in \mathbb{R}$ are the channel gains from Relay 1 to Destination 2 and Relay 2 to Destination 1, respectively.  Without loss of generality, we take $a \neq 0$ and $b \neq 0$. The case of $a=0$ or $b=0$ follows from continuity. The transmitted signals at the two relays must satisfy the average power constraints: for any $x_k^n$ that Relay $k$ sends into the channel, it must satisfy
\begin{align}
\frac{1}{n}\sum_{i=1}^n x_{ki}^2\le P_k,\quad k=1,2. \nonumber
\end{align}
%Without of loss of generality, we may assume $a>0, b \in \mathbb{R}$. Since the case of $a<0$ can easily be transformed into the case of $a>0$ by defining $X_2'=-X_2$ and $Y_2'=-Y_2$. This results in the following equivalent channel due to the one-to-one mapping
%\begin{align}
%Y_1&=X_1-a X_2'+U_1, \label{equivalent01}\\
%Y_2'&=-b X_1+X_2'+U_2. \label{equivalent02}
%\end{align}
%where $-a$ is positive. 

\section{An Upper Bound On the sum capacity of the 2-user Gaussian multiple access diamond channel}
The following of this paper finds an upper bound on the sum capacity of the 2-user Gaussian multiple access diamond channel.

%For $\bar{\rho} \in [-1,1]$ and $\nu \in [1, \infty)$, define 
%\begin{align}
%C_{\text{MIMO}12}^{\nu}(\bar{\rho})& \triangleq \max_{(R_1, R_2) \in \mathcal{C}_{\text{MIMO}}(\bar{\rho})} \quad R_1+\nu R_2, \nonumber\\
%C_{\text{MIMO}21}^{\nu}(\bar{\rho})& \triangleq \max_{(R_1, R_2) \in \mathcal{C}_{\text{MIMO}}(\bar{\rho})} \quad R_2+\nu R_1 ,\nonumber
%\end{align}
For $\bar{\rho} \in [-1,1]$, define 
\begin{align}
C_{\text{MIMO}}^{\text{sum}}(\bar{\rho})& \triangleq \max_{(R_1, R_2) \in \mathcal{C}_{\text{MIMO}}(\bar{\rho})} \quad R_1+ R_2, \label{MIMOsum}
\end{align}
where $\mathcal{C}_{\text{MIMO}}(\bar{\rho})$ denotes the capacity region of the broadcast channel described in (\ref{GC01}) and (\ref{GC02}) where $\mathbf{X} \triangleq \begin{bmatrix} X_1 & X_2 \end{bmatrix}^T$ is the transmitted signal of the 2 antennas of the transmitter, and $Y_1$ and $Y_2$ are the received signals of the single-antenna Receivers 1 and 2, respectively. The input of the transmitter must satisfy a covariance constraint, i.e.,
\begin{align}
E \left[ \mathbf{X} \mathbf{X}^T \right] \preceq \begin{bmatrix} P_1 & \bar{\rho}\sqrt{P_1P_2}\\  \bar{\rho}\sqrt{P_1P_2} & P_2 \end{bmatrix}. \nonumber
\end{align}
The  capacity region of the MIMO broadcast channel, i.e.,  $\mathcal{C}_{\text{MIMO}}(\bar{\rho})$, has been found in \cite{H.Weingarten:2006, T.Liu:2007, C.Nair:2014}. $C_{\text{MIMO}}^{\text{sum}}(\bar{\rho})$ defined in (\ref{MIMOsum}) is the sum capacity of the corresponding MIMO broadcast channel and it has been found in \cite{Viswanath:2003, Vishwanath:2003, Yu:BC2004}. 

Before we introduce the main theorem, let us define the following functions for $\rho \in [-1,1]$, 
\begin{align}
f_A(\rho) &\triangleq C_1+\frac{1}{2} \log \left(1+\max\{a^2,1\} (1-{\rho}^2) P_2 \right), \nonumber\\
f_B(\rho) &\triangleq C_2+\frac{1}{2} \log \left(1+\max\{b^2,1\} (1-{\rho}^2) P_1 \right), \nonumber\\
f_C(\rho) &\triangleq C_1+C_2-\frac{1}{2} \log \frac{1}{1-\rho^2}, \nonumber 
%f_C^m(\rho) &\triangleq \frac{1}{2}[C_1+C_2-\frac{1}{2} \log \frac{1}{1-\rho^2}+C_{\text{MIMOm}}^1(\rho)] \quad m=12, 21
%f_D(\rho) & \triangleq C_{\text{MIMO}12}^1 (\rho) \nonumber\\
\end{align}
and the following variables
\begin{align}
\rho_x=\sgn(x) \left(\sqrt{1+\frac{1}{4x^2P_1P_2}}-\sqrt{\frac{1}{4x^2P_1P_2}} \right), \quad x=a,b, \nonumber
%\\
%\rho_b=\sgn(b) \left(\sqrt{1+\frac{1}{4b^2P_1P_2}}-\sqrt{\frac{1}{4b^2P_1P_2}} \right). \nonumber
\end{align}
where $\sgn(\cdot)$ is the sign function of $\cdot$, and the following sets
\begin{align}
\mathcal{A}_x=
\left\{
\begin{array}{ll}
{[0,\rho_x]} &  \text{ if } x \geq 0 \\
{[\rho_x,0]} & \text{ if } x <0 
\end{array}
\right., \quad x=a,b. \nonumber 
\end{align}

The following is the main result of this paper.
\begin{Theo}\label{ub2}
The sum-rate $R_1+R_2$ is achievable for the 2-user Gaussian multiple access diamond channel only if it satisfies
\begin{align}
R_1+R_2 &\leq \max_{\rho \in [-1,1]} \min \left\{f_A(\rho),f_B(\rho),f_C(0),C_{\text{MIMO}}^{\text{sum}} (\rho) \right\},
  \text{ and }\label{Theo0102}\\
R_1+R_2 &\leq \max_{\rho \in \mathcal{A}_x} \min \left\{f_A(\rho),f_B(\rho), f_C(0), C_{\text{MIMO}}^{\text{sum}}(\rho), \frac{1}{2} \left(f_C(\rho)+C_{\text{MIMO}}^{\text{sum}}(\rho) \right) \right\}, \label{Theo0101}
%R_1+R_2 &\leq \max \left\{\max_{\rho \in \mathcal{A}_a} \min \{f_A(\rho),f_B(\rho),f_C^{21}(\rho),C_{\text{MIMO}21}^1(\rho)
% \} \right\}  \text{ and }\label{Theo0103}\\
%R_1+R_2 &\leq \max \left\{\max_{\rho \in [-1,1]\bigcap\mathcal{A}_a^c} \min \{f_A(\rho),f_B(\rho),f_C(0),C_{\text{MIMO}21}^1(\rho)
% \} \right\}  \text{ and }\label{Theo0104}
\end{align}
for both $x=a$ and $x=b$. 
\end{Theo}
\begin{IEEEproof}
The proof is in Appendix \ref{proof01}.
\end{IEEEproof}

\begin{figure}[t!]
\centering
\includegraphics[width=3in]{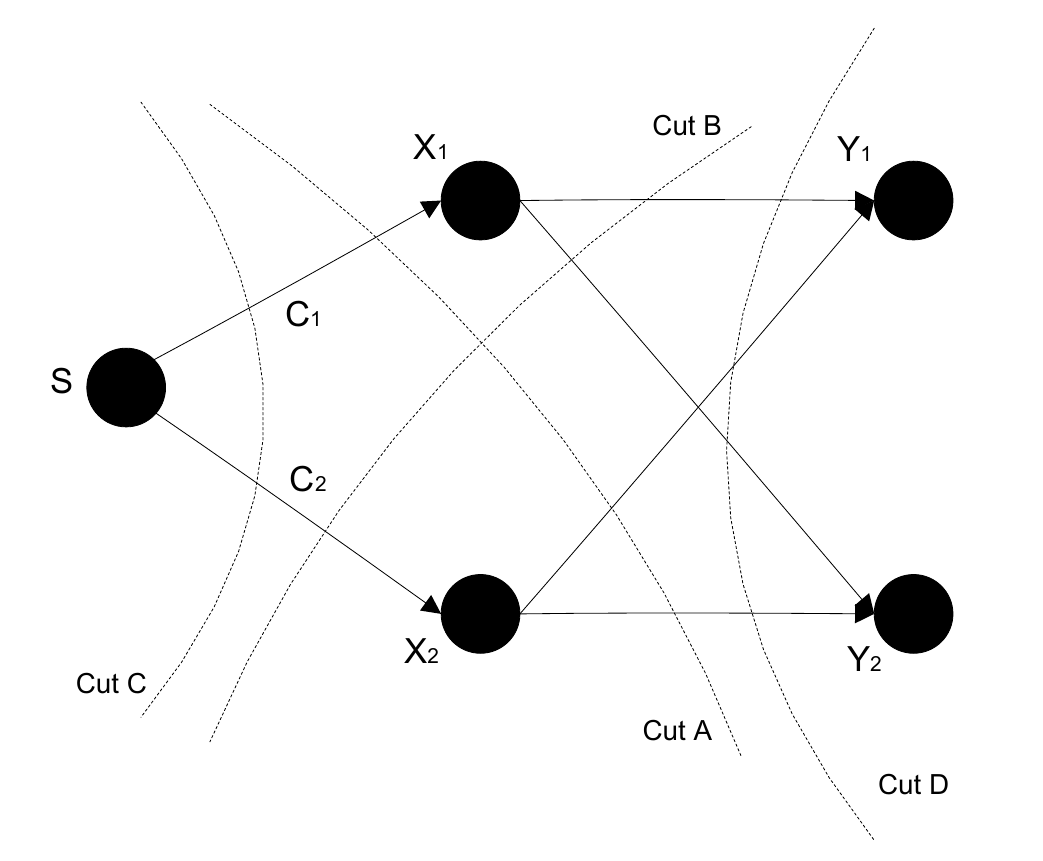}
\caption{Cut-set bounds for the channel} \label{csbound_pic}
\end{figure}

In Thoerem \ref{ub2}, (\ref{Theo0102}) is proved using the cut-set bound from the four cuts, i.e., Cuts A, B, C and D of Fig. \ref{csbound_pic}, on the sum rate $R_1+R_2$. The more difficult part is to prove that when $\rho$ satisfies $\rho \in \mathcal{A}_x$, $x=a,b$, then we have (\ref{Theo0101}), which is strictly tighter than (\ref{Theo0102}). The converse techniques we use to prove this include
1) the bounding of the correlation between the transmitted signals of the two relays via an auxiliary random variable 
\cite{W.Kang:2015, G.Kramer:2014, G.Kramer:2015, G.Kramer:2016}, which was inspired by Ozarow in solving the Gaussian multiple description problem \cite{L.Ozarow:1980};
2) the single-letterization technique from \cite[page 314, equation (3.34)]{Csiszar:book};
3) the entropy power inequality (EPI) \cite[Lemma I]{Bergmans:1974}; and
4) the derivation of the capacity region of the Gaussian MIMO broadcast channel with private messages in \cite[Section III.A]{C.Nair:2014}. 

The existing simple cut-set upper bound on the sum capacity is
\begin{align}
R_1+R_2 \leq \min \{f_C(0), \max_{\rho \in [-1,1]}C_{\text{MIMO}}^{\text{sum}}(\rho)\}, \label{simpleupper}
\end{align}
which is the minimum of the capacity of Cuts C and D of Fig. \ref{csbound_pic}.  Comparing this with the result of Theorem \ref{ub2}, we see that the upper bound of (\ref{Theo0102}) is tighter than the existing simple cut-set bound, as it further considers the capacities of Cuts A and B. Moreover, we have the upper bound in (\ref{Theo0101}), which is strictly tighter than the cut-set bound in (\ref{Theo0102}) when $\rho$ satisfies $\rho \in \mathcal{A}_x$, for $x=a,b$. Thus, Theorem \ref{ub2} provides a novel upper bound that is tighter than the existing simple cut-set bound of (\ref{simpleupper}). 

%%%%%%%%%%%%%%%%%%%%%%%%%%%%%%%%%%%%%%%%%%%%
%%            the extension
%%%%%%%%%%%%%%%%%%%%%%%%%%%%%%%%%%%%%%%%%%%%

\section{Numerical Results}
To illustrate the tightness of the derived upper bound in Theorem \ref{ub2}, we plot and compare the existing simple cut-set upper bound on the sum capcity in (\ref{simpleupper}), the new cut-set upper bound of (\ref{Theo0102}), the new upper bound of Theorem \ref{ub2}, and the achievable sum rates of existing schemes for the 2-user Gaussian multiple access diamond channel. 

\begin{figure}[t!]
\centering
\includegraphics[width=5.5in]{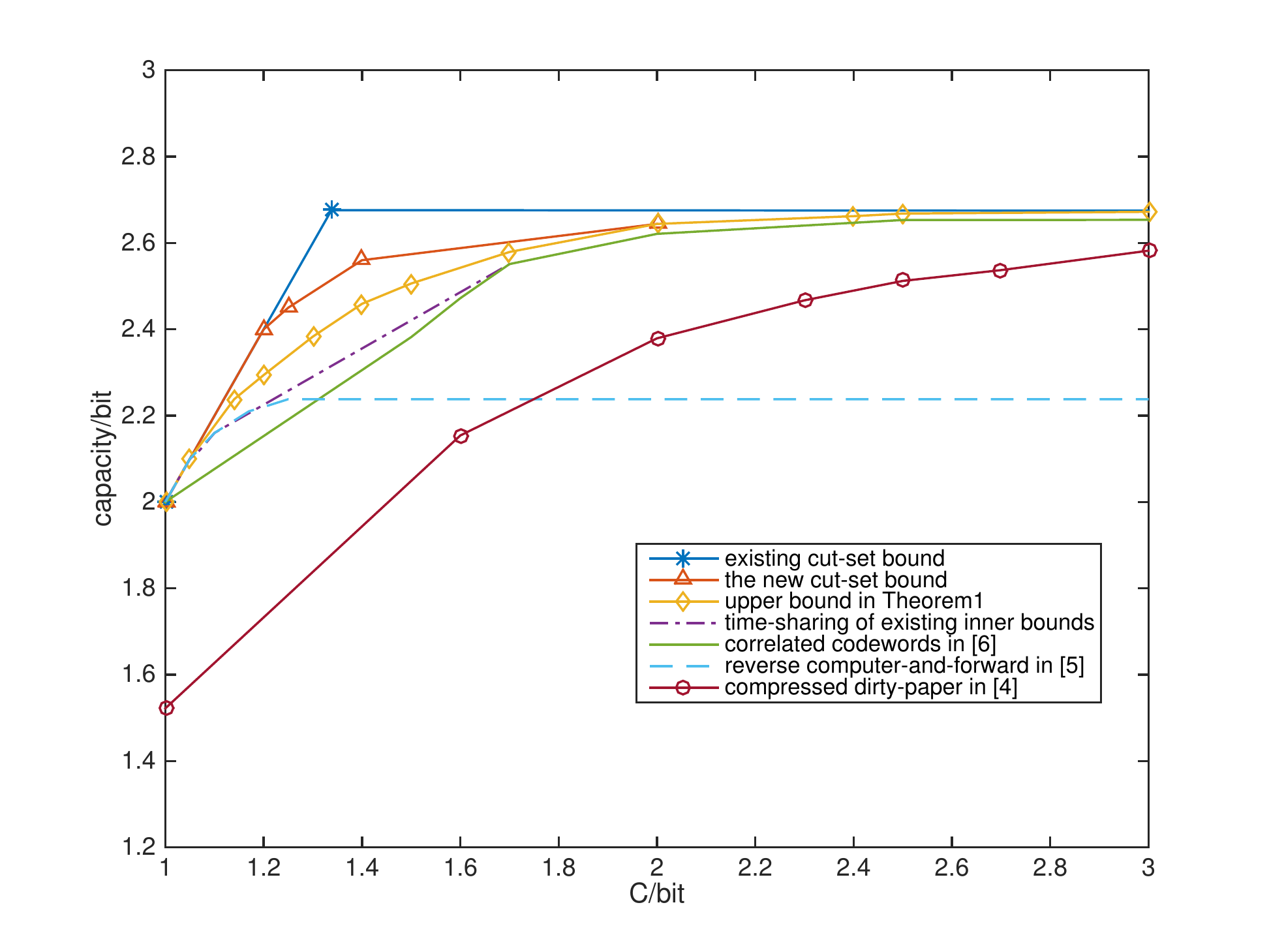}
\caption{Upper and lower bounds on the sum capacity for the case of  $a=b=0.9$, $P_1=P_2=10$ and $C_1=C_2=C$.} \label{simulation_pic1}
\end{figure}

The results are shown in Fig. \ref{simulation_pic1} for the symmetric case of $a=b=0.9$, $P_1=P_2=10$ and $C_1=C_2=C$. We only plot the region of $C \in [1,3]$, since this is the interesting case where the existing simple cut-set upper bound and the existing lower bounds on the sum capacity do not meet. 
%As can be seen, the existing cut-set upper bound is denoted by the stared line, the new cut-set bound of (\ref{Theo0102}) is denoted by the triangle line, and the upper bound derived in Theorem \ref{ub2} is denoted by the diamond line. 
As can be seen, in the region of $C \in  [1.2, 2.55]$, the the new cut-set bound of (\ref{Theo0102}) improves upon the existing simple cut-set bound of (\ref{simpleupper}), which means that in this region, it is beneficial to consider the cross-cuts in the cut-set bound, i.e., Cuts A and B. In the region of $C \in [1.05, 2]$, the upper bound of Theorem \ref{ub2} improves upon the new cut-set bound of (\ref{Theo0102}), which means that in this region, the upper bound (\ref{Theo0101}) is strictly tighter. Overall, 
in the region of $C \in [1.05, 2.55]$, our new upper bound improves upon the existing simple cut-set upper bound strictly. Furthermore, in the region of $C \in [1.05, 2]$, the improvement is rather significant. 

The sum rate achieved by the achievable schemes of sending correlated codewords by the relays \cite{N.Liu:2014} , the compressed dirty-paper coding allowing correlated quantization noise \cite{S-HPark:2013} and the reverse computer-and-forward scheme \cite{S-NHong:2013} are denoted by the solid, circled, and dashed lines, respectively. Furthermore, the sum rate of the time-sharing of all the existing achievable schemes, which is the largest known lower bound for the sum capacity, is denoted by the dot-dashed line. In the gap between the derived upper bound in Theorem \ref{ub2}, i.e., the diamond line, and the largest known lower bound for the sum capacity, i.e., the   dot-dashed line, lies the sum capacity of the 2-user Gaussian multiple access diamond channel for this symmetric case, and as we can see, the gap is not large, which means that the existing achievable schemes perform reasonably well for this scenario. 

\begin{figure}[t!]
\centering
\includegraphics[width=5.5in]{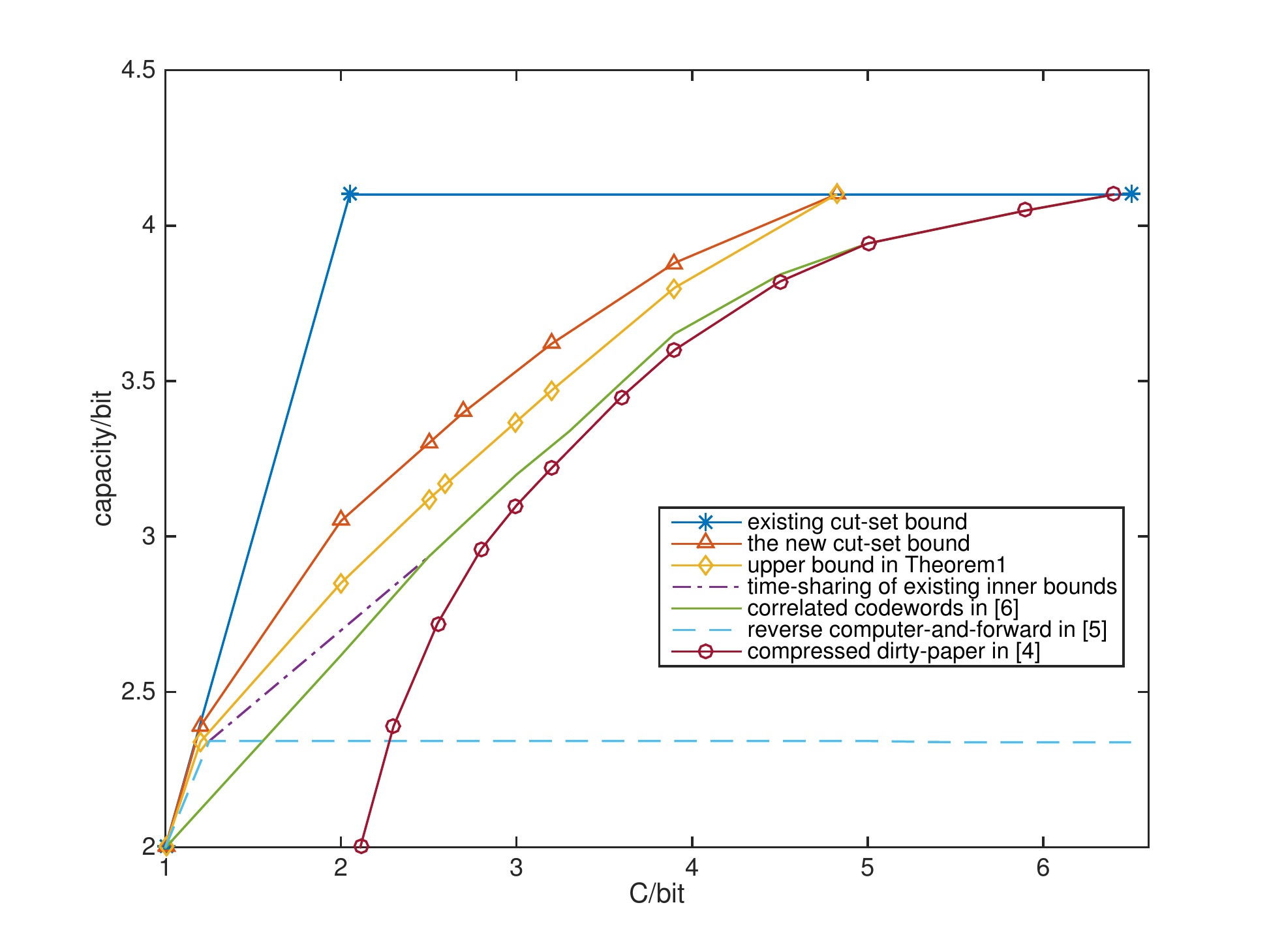}
\caption{Upper and lower bounds on the sum capacity for the case of  $a=0.9$, $b=-0.9$, $P_1=P_2=10$ and $C_1=C_2=C$.} \label{simulation_pic2}
\end{figure}

In the case of $a=0.9$, $b=-0.9$, $P_1=P_2=10$ and $C_1=C_2=C$, the results are shown in Fig. \ref{simulation_pic2}, and similar observations as Fig. \ref{simulation_pic1} follow.

\section{Conclusion}
In this paper, we derive a novel upper bound on the sum capacity of the 2-user Gaussian multiple access diamond channel. This is done by utilizing the converse tools of the multiple access diamond channel and that of the Gaussian MIMO broadcast channel. Through numerical results, we show that the derived upper bound improves upon the existing simple cut-set upper bound significantly, and as a result, the gap between the lower and upper bounds on the sum capacity is greatly reduced when the capacities of the backhaul links are in the medium range.

\appendices
\section{proof of Theorem 1} \label{proof01}
For any sequence of $(2^{nR_1},2^{nR_2},n,\epsilon_n)$ code, let $X_k^n$ denote the input of Relay $k$ into the $n$ uses of the channel $p(y_1,y_2|x_1,x_2)$, and $Y_k^n$ denote the corresponding output received at Receiver $k$, $k=1,2$. Due to the power constraint, we have
\begin{align}
\frac{1}{n} \sum_{i=1}^n E[X_{ki}^2] \leq P_k, \quad k=1,2. \label{PC02}
\end{align}
Define of a random variable $Q$ that is independent of everything else and uniformly distributed on $\{1,2, \cdots, n\}$, further define
\begin{align}
X_1 \triangleq X_{1Q}, X_2 \triangleq X_{2Q}, Y_1 \triangleq Y_{1Q}, Y_2 \triangleq Y_{2Q}. \label{XQ}
\end{align}
Define 
%$\mathbf{X}=\begin{bmatrix} X_1 & X_2\end{bmatrix}^T$, and
the
correlation coefficient between $X_1$ and $X_2$ as
\begin{align}
&\rho \triangleq \frac{E[X_1X_2]}{\sqrt{E[X_1^2]E[X_2^2]}}. \nonumber
\end{align}
Note that $\rho \in [-1,1]$. Further define
\begin{align}
\bar{P}_k \triangleq E[X_k^2], \quad k=1,2. \nonumber
\end{align}
From (\ref{PC02}) and (\ref{XQ}), we have
\begin{align}
\bar{P}_k=E[X_k^2]=E[X_{kQ}^2]=E_Q \left[ E_{X|Q} \left[X_{kQ}^2|Q=i \right] \right]=\frac{1}{n} \sum_{i=1}^n E[X_{ki}^2] \leq P_k, \quad k=1,2. \label{PC}
\end{align}
Define
\begin{align}
\rho^* \triangleq \frac{ \sqrt{\bar{P}_1 \bar{P}_2}}{\sqrt{P_1 P_2}} \rho. \label{rho_star}
\end{align}
Based on (\ref{PC}), we have
\begin{align}
\left|\rho^* \right| \leq  \left|\rho \right|. \label{BigSmall}
\end{align}
Hence, $\rho^* \in [-1,1]$. 
Define $\mathbf{X} \triangleq \begin{bmatrix} X_1 & X_2 \end{bmatrix}^T$, and further define $\mathbf{K}$ as
\begin{align}
\mathbf{K} \triangleq \begin{bmatrix} P_1 & \rho^*\sqrt{P_1 P_2}\\  \rho^* \sqrt{P_1 P_2} & P_2 \end{bmatrix}. \nonumber
\end{align}
We can see that 
\begin{align}
E[\mathbf{X} \mathbf{X}^T] =\begin{bmatrix} \bar{P}_1 & \rho\sqrt{\bar{P}_1 \bar{P}_2}\\  \rho\sqrt{\bar{P}_1\bar{P}_2} & \bar{P}_2 \end{bmatrix} \preceq \begin{bmatrix} P_1 & \rho\sqrt{\bar{P}_1 \bar{P}_2}\\  \rho\sqrt{\bar{P}_1\bar{P}_2} & P_2 \end{bmatrix}=\begin{bmatrix} P_1 & \rho^*\sqrt{P_1 P_2}\\  \rho^* \sqrt{P_1P_2} & P_2 \end{bmatrix} =\mathbf{K}. \label{StarAdd}
\end{align}
%From the cut-set bound, we always have
%\begin{align}
%R_1+R_2 \le C_1+C_2 \nonumber
%\end{align}
%and
%\begin{align}
%R_1+R_2 \le C_{\text{MIMO}}(\rho) \nonumber 
%\end{align}
%for some $\rho \in [0,1]$. 
%where the right-hand side is the capacity of the multiple access diamond channel with two destinations
%assuming the ideal case of $C_1, C_2 \rightarrow \infty$.

Now, based on the four cuts demonstrated in Fig. \ref{csbound_pic}, we have the following cut-set upper bounds on the sum capacity, i.e., $R_1+R_2$:
\begin{enumerate}
\item Considering Cut C, we have 
\begin{align}
R_1+R_2 \leq C_1+C_2=f_C(0). \label{CutCut01}
\end{align}
\item Considering Cut D, due to (\ref{StarAdd}), we have 
\begin{align}
R_1+R_2 \leq C_{\text{MIMO}}^{\text{sum}}(\rho^*). \label{CutCut02}
\end{align} 
%{\color{red} do we need this cut, or can it be derived from other cuts?}
%\\
%{
%\color{blue}
%I think that the cut D can be deduced from the capacity of MIMO broadcast channel which is applied in the all other cuts. 
%So In the Fig.2, I have just drawn three cuts. 
%}
\item Considering Cut B, we have two cases:
\begin{enumerate}
\item For the case of $|b| \leq 1$, 
\begin{align}
&n (R_1+R_2)\nonumber\\
&=H(W_1,W_2) \nonumber\\
&=H(W_1, W_2, X_2^n) \label{add01} \\
&=H(X_2^n)+H(W_1, W_2|X_2^n) \label{add08}\\
& = H(X_2^n)+H(W_1|X_2^n, W_2)+H(W_2|X_2^n) \nonumber\\
& \leq n C_2+H(W_1|X_2^n, W_2)+H(W_2|X_2^n) \nonumber\\
& =n C_2+I(W_1;Y_1^n|X_2^n, W_2)+H(W_1|Y_1^n, X_2^n, W_2)+I(W_2;Y_2^n|X_2^n)+H(W_2|Y_2^n, X_2^n) \nonumber\\
&\leq n C_2+I(W_1;Y_1^n|X_2^n, W_2)+I(W_2;Y_2^n|X_2^n)+2 n \epsilon_n \label{FanoWai}\\
&= n C_2+I(X_1^n;Y_1^n|X_2^n, W_2)+I(W_2;Y_2^n|X_2^n)+2 n \epsilon_n, \label{add02}
\end{align}
where (\ref{add01}) follows from the fact that without loss of generality, we consider deterministic encoding at the source node, i.e., $(X_1^n, X_2^n)$ is a deterministic function of $(W_1, W_2)$, (\ref{FanoWai}) follows from Fano's inequality,  (\ref{add02}) follows from the fact that we consider deterministic encoders and the Markov Chain $W_1 \rightarrow (X_1^n, X_2^n, W_2) \rightarrow Y_1^n$. 
Define $\tilde{Y}_2^n$ as  the following channel
\begin{align}
\tilde{Y}_2^n=Y_1^n+\left(\frac{1}{b}-a \right)X_2^n+\tilde{U}^n, \nonumber
\end{align}
where $\tilde{U}^n$ is an i.i.d. sequence of Gaussian random variables with zero mean and variance $\frac{1}{b^2}-1$, and it is independent of everything else. Note that
given $X_2^n$, $\tilde{Y}_2^n$ is a physically degraded version of $Y_1^n$. Furthermore, note the similarity between
\begin{align}
\tilde{Y}_2^n&=X_1^n+\frac{1}{b} X_2^n+U_1^n+\tilde{U}^n \quad\text{and} \quad Y_2^n=b X_1^n+X_2^n+U_2^n, \nonumber
\end{align}
which means that we have 
\begin{align}
I(W_2;Y_2^n|X_2^n)=I(W_2;\tilde{Y}_2^n|X_2^n). \nonumber
\end{align}
Thus, from (\ref{add02}), we continue to write as follows while for the simplicity of presentation, we have dropped the $2 n \epsilon_n$ term, 
\begin{align}
&n(R_1+R_2)\nonumber\\
 & \leq n C_2+I(X_1^n;Y_1^n|X_2^n, W_2)+I(W_2;\tilde{Y}_2^n|X_2^n) \nonumber\\
 &=n C_2+\sum_{i=1}^n I(X_{1i};Y_{1i}|X_2^n, W_2, Y_1^{i-1})+I(W_2;\tilde{Y}_{2i}|X_2^n, \tilde{Y}_{2}^{i-1}) \nonumber\\
  &\leq n C_2+\sum_{i=1}^n I(X_{1i};Y_{1i}|X_2^n, W_2, \tilde{Y}_2^{i-1})+I(W_2;\tilde{Y}_{2i}|X_2^n, \tilde{Y}_{2}^{i-1}) \label{degradedNan}\\
& \leq n C_2+\sum_{i=1}^n I(X_{1i};Y_{1i}|X_2^n, W_2, \tilde{Y}_2^{i-1})+I(W_2, X_{2\{i\}^c}, \tilde{Y}_2^{i-1};\tilde{Y}_{2i}|X_{2i}) \nonumber
\end{align}
where (\ref{degradedNan}) follows from the fact that given $X_2^n$, $\tilde{Y}_2^n$ is a physically degraded version of $Y_1^n$.
Define auxiliary random variables 
\begin{align}
V_i=\left(W_2, X_{2\{i\}^c}, \tilde{Y}_2^{i-1} \right). \nonumber
\end{align}
Thus, we have
\begin{align}
n (R_1+R_2) 
&=n C_2+\sum_{i=1}^n \big(I(X_{1i};Y_{1i}|X_{2i}, V_i)+I(V_i;\tilde{Y}_{2i}|X_{2i}) \big) \label{add05}\\
&=n C_2+n \big(I(X_{1Q};Y_{1Q}|X_{2Q}, V_Q, Q)+I(V_Q;\tilde{Y}_{2Q}|X_{2Q}, Q) \big) \nonumber\\
& \leq n C_2+n \big(I(X_{1Q};Y_{1Q}|X_{2Q}, V_Q, Q)+I(V_Q,Q;\tilde{Y}_{2Q}|X_{2Q}) \big) \nonumber\\
& = n C_2+n \big(I(X_{1};Y_{1}|X_{2}, V)+I(V;\tilde{Y}_{2}|X_{2}) \big)  \label{add06}\\
& \leq n C_2+n E_{X_2} \left[I(X_{1};Y_{1}|X_{2}=x_2, V)+I(V;\tilde{Y}_{2}|X_{2}=x_2) \right] , \label{add07}
\end{align}
where (\ref{add05}) follows from the Markov Chain $V_i \rightarrow (X_{1i}, X_{2i}) \rightarrow Y_{1i}$, (\ref{add06}) follows from the definition in (\ref{XQ}) and 
\begin{align}
V \triangleq (V_Q, Q), \quad \tilde{Y}_2 \triangleq \tilde{Y}_{2Q}. \label{DefineLater}
\end{align}
Note that 
%$I(X_{1};Y_{1}|X_{2}=x_2, V)+I(V;\tilde{Y}_{2}|X_{2}=x_2)$ is 
the sum capacity of the degraded broadcast channel where the input of the channel is $X_1$ given $X_2=x_2$ and the outputs of the channel is $Y_1$ and $\tilde{Y}_2$, respectively, is given by \cite{Cover:book}
%Due to the input power constraint, i.e., $\text{Var}[X_1^2|X_2=x_2]=(1-\rho^2)\bar{P}_1$, 
%The sum capacity of degraded broadcast channel described above is
\begin{align}
%\frac{1}{2} \log \left(1+\text{Var}[X_1^2|X_2=x_2] \right)= 
&\max_{p(v,x_1)} I(X_{1};Y_{1}|X_{2}=x_2, V)+I(V;\tilde{Y}_{2}|X_{2}=x_2) \nonumber\\
& =\frac{1}{2} \log \left(1+E[(X_1-E[X_1|X_2=x_2])^2|X_2=x_2] \right).\nonumber
\end{align}
Hence, for the particular $p(v,x_1)$ as defined by the codebook and (\ref{DefineLater}), we have
\begin{align}
&I(X_{1};Y_{1}|X_{2}=x_2, V)+I(V;\tilde{Y}_{2}|X_{2}=x_2)  \nonumber\\
&\leq \frac{1}{2} \log \left(1+E[(X_1-E[X_1|X_2=x_2])^2|X_2=x_2] \right). \label{JensenLater}
\end{align}
Hence, following from (\ref{add07}) and (\ref{JensenLater}), we have
\begin{align}
R_1+R_2 &\leq  C_2+E_{X_2} \left[\frac{1}{2} \log \left(1+E[(X_1-E[X_1|X_2=x_2])^2|X_2=x_2] \right) \right] \nonumber\\
& \leq C_2+\frac{1}{2} \log \left(1+E_{X_2}\left[E[(X_1-E[X_1|X_2=x_2])^2|X_2=x_2] \right]\right) \label{WeiConvexity}\\
& \leq C_2+\frac{1}{2} \log \left(1+E[X_1^2]-\frac{E^2[X_1X_2]}{E[X_2^2]} \right) \label{MMSEagain}\\
&=C_2+\frac{1}{2} \log \left(1+(1-{\rho}^2) \bar{P}_1 \right) \nonumber\\
& \leq  C_2+\frac{1}{2} \log \left(1+(1-{\rho^*}^2) P_1 \right), \label{cutresult01}
\end{align}
where (\ref{WeiConvexity}) follows from the convexity of the $\log(\cdot)$ function, (\ref{MMSEagain}) follows from the fact that the mean-squared error (MSE) of the optimal Bayes least square (BLS) estimator is smaller than that of the linear least squared (LLS) estimator, and (\ref{cutresult01}) follows from (\ref{PC}) and (\ref{BigSmall}). 

\item Similarly, for the case of $|b| >1$, following from (\ref{add08}), we have
\begin{align}
&n (R_1+R_2)\nonumber\\
&=H(X_2^n)+H(W_1, W_2|X_2^n) \nonumber\\
& = H(X_2^n)+H(W_2|X_2^n, W_1)+H(W_1|X_2^n) \nonumber\\
& \leq  n C_2+I(X_1^n;Y_2^n|X_2^n, W_1)+I(W_1;Y_1^n|X_2^n)+2 n \epsilon_n. \nonumber
\end{align}
By following similar steps as (\ref{add02}) to (\ref{cutresult01}), we may conclude that 
\begin{align}
R_1+R_2 \leq  C_2+\frac{1}{2} \log \left(1+b^2 (1-{\rho^*}^2) P_1\right). \label{cutresult02}
\end{align}
Combining (\ref{cutresult01}) and (\ref{cutresult02}), we have
\begin{align}
R_1+R_2 \leq  C_2+\frac{1}{2} \log \left(1+\max\{b^2,1\} (1-{\rho^*}^2) P_1 \right)=f_B(\rho^*). \label{CutCut03}
\end{align}
\end{enumerate}
\item Due to symmetry, when we consider Cut A, we obtain
\begin{align}
R_1+R_2 \leq  C_1+\frac{1}{2} \log \left(1+\max\{a^2,1\} (1-{\rho^*}^2) P_2 \right)=f_A(\rho^*). \label{CutCut04}
\end{align}
\end{enumerate}

Thus, from (\ref{CutCut01}), (\ref{CutCut02}), (\ref{CutCut03}) and (\ref{CutCut04}), we have
\begin{align}
R_1+ R_2 
 \leq \min \left[f_A(\rho^*), f_B(\rho^*), f_C(0), C_{\text{MIMO}}^{\text{sum}}(\rho^*) \right]. \label{theother}
\end{align}
Note that (\ref{theother}) is valid for all values of $\rho^* \in [-1,1]$.

We now proceed to derive another upper bound on $R_1+ R_2$ which is valid when $\rho^*$ satisfies 
\begin{align}
0 < \rho^* \leq \sqrt{1+\frac{1}{4b^2P_1P_2}}-\sqrt{\frac{1}{4b^2P_1P_2}} \label{rhoCond}
\end{align}
in the case of $b>0$. If $b<0$, then the upper bound is valid if $\rho^*$ satisfies
\begin{align}
-\sqrt{1+\frac{1}{4b^2P_1P_2}}+\sqrt{\frac{1}{4b^2P_1P_2}} \leq  \rho^* <0. \label{rhoCond02}
\end{align}
Using Fano's inequality, we have
\begin{align}
n(R_1+R_2) &\leq I(W_1,W_2;Y_1^n, Y_2^n)+n \epsilon_n \nonumber\\
& \leq I(X_1^n, X_2^n;Y_1^n, Y_2^n)+n \epsilon_n \label{deterministic}\\
& \leq H(X_1^n, X_2^n)+n \epsilon_n \nonumber\\
&=H(X_1^n)+H(X_2^n)-I(X_1^n;X_2^n)+n \epsilon_n \nonumber\\
& \leq n C_1+n C_2-I(X_1^n;X_2^n)+n \epsilon_n, \label{CutSetSet}
\end{align}
where (\ref{deterministic}) follows from the Markov chain $(W_1, W_2) \rightarrow (X_1^n, X_2^n) \rightarrow (Y_1^n, Y_2^n)$.
%For the case of $\mu \geq 1$
%and $0 < \rho \leq \sqrt{1+\frac{1}{4b^2P_1P_2}}-\sqrt{\frac{1}{4b^2P_1P_2}}$, 
%we derive the following. The case of $\mu \leq 1$
%and $0<\rho \leq \sqrt{1+\frac{1}{4a^2P_1P_2}}-\sqrt{\frac{1}{4a^2P_1P_2}}$ 
%can be derived similarly by swapping the indices 1 and 2 and change the number $b$ to $a$. 
Using Fano's inequality, we further have
\begin{align}
n(R_1+R_2) &\le H(W_1,W_2)=H(W_2)+H(W_1|W_2) \nonumber \\
	&\le I(W_2;Y_2^n)+I(W_1;Y_1^n|W_2)+H(W_2|Y_2^n)+H(W_1|Y_1^n,W_2) \nonumber \\
	&\le I(W_2;Y_2^n)+I(W_1;Y_1^n|W_2)+2n\epsilon_n \nonumber \\
	&\le I(W_2;Y_2^n)+I(X_1^n,X_2^n;Y_1^n|W_2)+2n\epsilon_n, \nonumber \\
%\end{align}
%and 
%\begin{align}
%nR_2 &= H(W_2) = I(W_2;Y_2^n)+H(W_2|Y_2^n) \nonumber \\
%	&\le I(W_2;Y_2^n)+n\epsilon_n. \label{102}
%\end{align}
%For notational simplicity, from now on, we omit the $\epsilon_n$ term which will go to zero and $n \rightarrow \infty$. 
%we have:
%\begin{align}
%&n(R_1+R_2) \nonumber \\
%&\quad \leq I(X_1^n,X_2^n;Y_1^n|W_2)+I(W_2;Y_2^n)  \label{103}\\
&\le I(X_1^n,X_2^n;Y_1^n|W_2)+I(X_1^n, X_2^n;Y_2^n)- I(X_1^n, X_2^n;Y_2^n|W_2)+2n\epsilon_n, \label{104}
\end{align}
where 
%(\ref{103}) follows from (\ref{lem4eqeq1}) and (\ref{102}) and the facts that $\lambda \leq 1$ and $\mu \geq 1$, and 
(\ref{104}) is because of $W_2\rightarrow(X_1^n,X_2^n)\rightarrow(Y_1^n,Y_2^n)$ forms a Markov chain. %and the $\sup$ is over all distribution $p(v|x_1^n, x_2^n)$ and the mutual informations are evaluated using the distribution
%\begin{align}
%p(v,x_1^n,x_2^n, y_1^n, y_2^n)=p(x_1^n,x_2^n)p(v|x_1^n,x_2^n)p(y_1^n,y_2^n|x_1^n,x_2^n), \label{MCMC2}
%\end{align}
%and (\ref{004}) is because the $\sup$ is over $V$ that satisfies the Markov Chain $V\rightarrow(X_1^n,X_2^n)\rightarrow Y_2^n$ as indicated by (\ref{MCMC2}).
Thus, omitting the $\epsilon_n$ term which will go to zero and $n \rightarrow \infty$, from (\ref{CutSetSet}) and (\ref{104}), we have
\begin{align}
&2n(R_1+ R_2)  \nonumber\\
	%& \le (nC_1+nC_2-I(X_1^n;X_2^n))+nR_1+nR_2 \label{105} \\
	& \le n(C_1+C_2)- I(X_1^n;X_2^n)+I(X_1^n,X_2^n;Y_2^n)+I(X_1^n,X_2^n;Y_1^n|W_2)\nonumber\\
	&\quad-I(X_1^n,X_2^n;Y_2^n|W_2) \nonumber\\
	&\le n(C_1+C_2)+I(X_1^n,X_2^n;Y_2^n)-I(X_1^n,X_2^n;Z^n) + I(X_1^n;Z^n|X_2^n) \nonumber \\
	&\quad+ I(X_2^n;Z^n|X_1^n)+ I(X_1^n,X_2^n;Y_1^n|W_2)- I(X_1^n,X_2^n;Y_2^n|W_2), \label{Theo1ieq4}	
\end{align}
where 
%(\ref{105}) follows and the fact that $\lambda \geq 0$, 
%(\ref{Theo1ieq3another}) follows from (\ref{104}), and 
(\ref{Theo1ieq4}) follows by introducing a sequence of auxiliary random variables $Z^n$ and utilizing the fact that
\begin{align}
I(X_1^n;X_2^n) &=I(X_1^n;Z^n)-I(X_1^n;Z^n|X_2^n)+I(X_1^n;X_2^n|Z^n) \nonumber \\
			&\ge I(X_1^n;Z^n)-I(X_1^n;Z^n|X_2^n) \nonumber \\
			&=I(X_1^n,X_2^n;Z^n)-I(X_2^n;Z^n|X_1^n)-I(X_1^n;Z^n|X_2^n). \nonumber
\end{align}
The above derivation is true for any $Z^n$. 

Next, we perform the single-letterization of (\ref{Theo1ieq4}). To do this, we restrict ourselves to consider $Z^n$ that is the output of the following memoryless Gaussian channel with $Y_2^n$ being the input: 
\begin{align}
Z=Y_2+U_3, \label{014}
\end{align}
where $U_{3}$ is a Gaussian random variable with zero mean and variance $N_3$. 
Further define
$
Z \triangleq Z_Q. \nonumber
$
We single-letterize (\ref{Theo1ieq4}) by single-letterizing each of the following three terms. 
\begin{enumerate}
%%% ITEM 1
\item $I(X_1^n,X_2^n;Y_2^n)-I(X_1^n,X_2^n;Z^n)$

We have
\begin{align}
&I(X_1^n,X_2^n;Y_2^n)-I(X_1^n,X_2^n;Z^n)\nonumber\\ &\le I(X_1^n,X_2^n;Y_2^n,Z^n)-I(X_1^n,X_2^n;Z^n) \nonumber \\
	&= I(X_1^n,X_2^n;Y_2^n|Z^n) \nonumber\\
	& =\sum_{i=1}^n I(X_1^n, X_2^n;Y_{2i}|Y_2^{i-1}, Z^n) \nonumber\\
	& \leq \sum_{i=1}^n I(X_{1i}, X_{2i};Y_{2i}|Z_i) \label{MCYZ}\\
%	&=\sum_{i=1}^n I(X_{1i}, X_{2i};Y_{2i}, Z_i)-I(X_{1i}, X_{2i}; Z_i)\nonumber\\
%	&=\sum_{i=1}^n I(X_{1i}, X_{2i};Y_{2i})+I(X_{1i}, X_{2i};Z_i|Y_{2i})-I(X_{1i}, X_{2i};Z_i) \nonumber\\
%	&=\sum_{i=1}^n I(X_{1i}, X_{2i};Y_{2i})-I(X_{1i}, X_{2i};Z_i) \label{012}\\
%	&=n \left[ I(X_1,X_2|Y_2, Q)-I(X_1,X_2;Z|Q) \right]
& =n I(X_1, X_2;Y_2|Z, Q) \nonumber\\
& \leq n I(X_1, X_2;Y_2|Z), \label{117}
\end{align}
where (\ref{MCYZ}) follows from the Markov chain $(X_1^n, X_2^n, Z^n, Y_2^{i-1}) \rightarrow (X_{1i}, X_{2i}, Z_i) \rightarrow Y_{2i}$, and (\ref{117}) follows from conditioning reduces entropy and the Markov chain $Q \rightarrow (Z, X_1, X_2) \rightarrow Y_2$.  
%where (\ref{012}) follows from the definition of $Z_i$ in (\ref{013}), which means that the  Markov chain  $ (X_{1i}, X_{2i}) \rightarrow Y_{2i} \rightarrow Z_i$ holds. 

%%% ITEM 2
\item $I(X_1^n;Z^n|X_2^n) +I(X_2^n;Z^n|X_1^n)$

%From the definition of $Z^n$ in (\ref{013}), we see that the following Markov chain holds.
%\begin{align}
%(X_1^n, X_2^n, Z^{i-1}) \rightarrow (X_{1i}, X_{2i}) \rightarrow Z_i, \quad \forall i=1,2,\cdots, n
%\end{align}
Based on (\ref{014}), we have
\begin{align}
I(X_1^n;Z^n|X_2^n) +I(X_2^n;Z^n|X_1^n) &\leq \sum_{i=1}^n I(X_{1i};Z_i|X_{2i})+I(X_{2i};Z_i|X_{1i}) \label{MCZ}\\
&=n (I(X_1;Z|X_2, Q)+I(X_2;Z|X_1,Q) ) \nonumber\\
& \leq n (I(X_1;Z|X_2)+I(X_2;Z|X_1)), \label{118}
\end{align}
where (\ref{MCZ}) follows from the Markov chain $(X_1^n, X_2^n, Z^{i-1}) \rightarrow (X_{1i}, X_{2i}) \rightarrow Z_i$. 

%%%%%% ITEM 3
\item $I(X_1^n,X_2^n;Y_1^n|W_2)-I(X_1^n,X_2^n;Y_2^n|W_2)$

From \cite[page 314, equation (3.34)]{Csiszar:book}, we have
\begin{align}
H(Y_1^n|W_2)-H(Y_2^n|W_2)=\sum_{i=1}^n \left[H(Y_{1i}|W_2, Y_{1}^{i-1}, Y_{2(i+1)}^n)-H(Y_{2i}|W_2, Y_{1}^{i-1}, Y_{2(i+1)}^n) \right]. \nonumber
\end{align}
Let us define $T_i \triangleq \left(W_2, Y_{1}^{i-1}, Y_{2(i+1)}^n \right)$, and further define
\begin{align}
T \triangleq T_Q. \label{defineV}
\end{align}
Note that the auxiliary random variables thus defined satisfy
\begin{align}
T \rightarrow (X_1, X_2) \rightarrow (Y_1, Y_2, Z). \label{110}
\end{align}
Based on the definition of the random variables in (\ref{XQ}) and (\ref{defineV}), we have
\begin{align}
&H(Y_1^n|W_2)-H(Y_2^n|W_2)= n \left(H(Y_{1}|T, Q)-H(Y_{2}|T, Q) \right), \text{ and } \nonumber\\
&n H(Y_1|T,Q)= \sum_{i=1}^n H(Y_{1i}|W_2, Y_{1}^{i-1}, Y_{2(i+1)}^n) \leq  \sum_{i=1}^n H(Y_{1i}|W_2, Y_{1}^{i-1})= H(Y_1^n|W_2). \nonumber
\end{align}
Thus, there exists a $\gamma \geq 0$ such that
\begin{align}
H(Y_1^n|W_2)=n \left(H(Y_1|T,Q)+\gamma \right), \quad H(Y_2^n|W_2)=n \left(H(Y_2|T, Q)+\gamma \right). \label{108}
\end{align}
Thus, we have
\begin{align}
&I(X_1^n,X_2^n;Y_1^n|W_2)-I(X_1^n,X_2^n;Y_2^n|W_2) \nonumber\\
=& H(Y_1^n|W_2)-H(Y_1^n|W_2, X_1^n, X_2^n)-H(Y_2^n|W_2)+ H(Y_2^n|X_1^n, X_2^n, W_2) \nonumber\\
=& n \bigg[H(Y_1|T, Q)+\gamma-H(Y_1|X_1, X_2,Q)- H(Y_2|T,Q)-\gamma + H(Y_2|X_1, X_2,Q) \bigg] \label{007}\\
=& n \bigg[H(Y_1|T,Q)-H(Y_1|X_1, X_2, T,Q)- H(Y_2|T,Q)+H(Y_2|X_1, X_2, T,Q)  \bigg]\label{109}\\
=& n \left[I(X_1,X_2;Y_1|T,Q)- I(X_1,X_2;Y_2|T,Q)  \right]\nonumber\\
=&n \left[I(X_1,X_2;Y_1|U)-I(X_1,X_2;Y_2|U)  \right], \label{UQ}
\end{align}
where (\ref{007}) follows from (\ref{108}), (\ref{109}) follows from (\ref{110}), and (\ref{UQ}) follows from the definition of the auxiliary random variable 
\begin{align}
U \triangleq (T,Q). \label{UQ1}
\end{align}
\end{enumerate}

From (\ref{Theo1ieq4}), (\ref{117}), (\ref{118}), (\ref{110}), (\ref{UQ}), we obtain the following single-letterization:
\begin{align}
2(R_1+ R_2) &\le C_1+C_2+ I(X_1,X_2;Y_2|Z)+ I(X_1;Z|X_2)+ I(X_2;Z|X_1)\nonumber \\
&\quad +I(X_1,X_2;Y_1|U)-I(X_1,X_2;Y_2|U), \label{Theo1mieq1}
\end{align}
where the mutual informations are evaluated using the joint distribution of the defined random variables $(X_1,X_2,Y_1,Y_2,Z,U)$ which satisfies
\begin{align}
p(x_1, x_2, y_1, y_2, z, u)=p(x_1, x_2)p(u|x_1,x_2)p(y_1,y_2|x_1,x_2)p(z|y_2). \label{disdis}
\end{align}
%where $p(y_1,y_2|x_1,x_2)$ refers to the channel in (\ref{GC01}) and (\ref{GC02}), and $p(z|y_2)$ refers to the channel in (\ref{014}). 
%for a fixed $p(z|x_1,x_2,y_1,y_2)$ and a fixed $p(y_1,y_2|x_1,x_2)$, (\ref{Theo2mieq1}) is concave for $p(x_1,x_2)$.

Next, we further derive an upper bound on (\ref{Theo1mieq1}) by using the fact that $p(y_1,y_2|x_1,x_2)$ in (\ref{disdis}), which refers to the channel in (\ref{GC01}) and (\ref{GC02}), and $p(z|y_2)$ in (\ref{disdis}), which refers to the channel in (\ref{014}), are Gaussian channels. 
%Further, we take the variance of the Gaussian noise in channel (\ref{014}) to be
%\begin{align}
%N_3 = \sqrt{b^2P_1P_2}(\frac{1}{\rho}-\rho)-1. \label{defineN}
%\end{align}
%Note that the condition $0 < \rho \leq \sqrt{1+\frac{1}{4b^2P_1P_2}}-\sqrt{\frac{1}{4b^2P_1P_2}}$ imply $N_3 \geq 0$. 
To derive an upper bound on (\ref{Theo1mieq1}), we provide an upper bound for the following three terms.
\begin{enumerate}
\item $I(X_1,X_2;Y_2|Z)$

We have
%Based on , we have $I(X_1,X_2;Y_2|Z)=I(X_1,X_2;Y_2)-I(X_1,X_2;Z)$, and therefore, we have
\begin{align}
&I(X_1,X_2;Y_2|Z) \nonumber\\
&=I(X_1,X_2;Y_2)-I(X_1,X_2;Z) \label{SaveLater}\\
&= \left(H(Y_2)- H(Z) \right)-\frac{1}{2} \log (2 \pi e) + \frac{1}{2} \log (2 \pi e)(1+N_3)  \nonumber\\
& \leq \frac{1}{2}  \log \left(b^2 P_1+P_2+2 b \rho^* \sqrt{P_1 P_2}+1\right)- \frac{1}{2} \log \left(\frac{b^2 P_1+P_2+2 b \rho^* \sqrt{P_1 P_2}+1+N_3}{1+N_3} \right), \label{LiuTie}
\end{align} 
where (\ref{SaveLater}) follows from the distribution of (\ref{disdis}), (\ref{LiuTie}) follows from the EPI \cite[Lemma I]{Bergmans:1974} and 
\begin{align}
E[(b X_1+X_2)^2]&=b^2 E[X_1^2]+E[X_2^2]+2 b \rho \sqrt{E[X_1^2] E[X_2^2]} \nonumber\\
& \leq b^2 P_1+P_2+2 b \rho \sqrt{\bar{P}_1 \bar{P}_2} \label{cov_scalar} \\
&= b^2 P_1+P_2+2 b \rho^* \sqrt{P_1 P_2}, \label{starstar}
\end{align}
where in (\ref{cov_scalar}), we have used (\ref{PC}), and (\ref{starstar}) follows from the definition of $\rho^*$ in (\ref{rho_star}). 
% and the fact the $\rho>0$. 

\item $ I(X_1;Z|X_2)$

Let us first calculate 
\begin{align}
h(Z|X_2=x_2) &\leq \frac{1}{2} \log (2 \pi e) E[(Z-E[Z|X_2=x_2])^2|X_2=x_2] \label{rig01}\\
&\leq \frac{1}{2} \log (2 \pi e) \left(E[Z^2]-\frac{E^2[ZX_2]}{E[X_2^2]}  \right)\label{rig02}\\
&=\frac{1}{2} \log (2 \pi e) (b^2(1-\rho^2)E[X_1^2]+1+N_3) \nonumber\\
& \leq \frac{1}{2} \log (2 \pi e) (b^2(1-\rho^2) P_1+1+N_3) \label{rig04}\\
& \leq \frac{1}{2} \log (2 \pi e) (b^2(1-\rho^{*2}) P_1+1+N_3), \label{star01}
\end{align}
where (\ref{rig01}) follows from the fact that given the covariance, the Gaussian distribution maximizes the differential entropy, (\ref{rig02}) follows from the fact that the MSE of the optimal BLS estimator is smaller than that of the LLS estimator, (\ref{rig04}) follows from (\ref{PC}), and (\ref{star01}) follows from (\ref{BigSmall}). Then, we have
\begin{align}
h(Z|X_2)&=\int_{\mathbb{R}} h(Z|X_2=x_2)f(x_2)d x_2 \nonumber\\
& \leq \frac{1}{2} \log (2 \pi e) (b^2(1-\rho^{*2})P_1+1+N_3), \label{rig03}
\end{align}
where (\ref{rig03}) follows from (\ref{star01}), and finally, we have
\begin{align}
I(X_1;Z|X_2) &=h(Z|X_2)-h(Z|X_1,X_2) \nonumber \\
	&\le \frac{1}{2}\log\frac{(1-\rho^{*2})b^2P_1+1+N_3}{1+N_3}. \label{Theo1ieq8}
\end{align}

\item $I(X_2;Z|X_1)$

Similarly to the calculation of $I(X_1;Z|X_2)$ above, we have:
\begin{align}
I(X_2;Z|X_1) &=h(Z|X_1)-h(Z|X_1,X_2) \nonumber \\
	&\le \frac{1}{2}\log\frac{(1-\rho^{*2})P_2+1+N_3}{1+N_3}. \label{Theo1ieq9}
\end{align}
\item $I(X_1,X_2;Y_1|U)-I(X_1,X_2;Y_2|U)$

We have
\begin{align}
&I(X_1,X_2;Y_1|U)-I(X_1,X_2;Y_2|U) \nonumber\\
& \leq \sup \limits_{p(u, x_1,x_2): E[\mathbf{X} \mathbf{X}^T] \leq \mathbf{K}} \left(I(X_1,X_2;Y_1|U)-I(X_1,X_2;Y_2|U) \right) \label{120}\\
& \leq C_{\text{MIMO}}^{\text{sum}}(\rho^*)- \max_{p(x_1,x_2): E[\mathbf{X} \mathbf{X}^T] \preceq \mathbf{K}} I(\mathbf{X};Y_2) \label{121}\\
%&=C_{\text{MIMO}}^{\frac{\mu-\lambda}{1-\lambda}}(\rho^*)-\frac{\mu-\lambda}{2(1-\lambda)} \log (b^2 %P_1+P_2+1+2\rho^* \sqrt{b^2 P_1 P_2}), \label{MIMO01}
&=C_{\text{MIMO}}^{\text{sum}}(\rho^*)-\frac{1}{2} \log (b^2 P_1+P_2+1+2b\rho^* \sqrt{P_1 P_2}), \label{MIMO01}
\end{align}
%where in (\ref{020}), we have defined $\mathbf{X}=\begin{bmatrix} X_1 & X_2 \end{bmatrix}^T$, and  $\bar{\mathbf{K}}$ is defined as $\bar{\mathbf{K}} \triangleq \begin{bmatrix} P_1 & \rho\sqrt{\bar{P}_1 \bar{P}_2}\\  \rho\sqrt{\bar{P}_1\bar{P}_2} & P_2 \end{bmatrix}$, 
where (\ref{120}) follows because $U$, $X_1$ and $X_2$ defined in (\ref{XQ}) and (\ref{UQ1}) satisfy the constraint of the optimization in (\ref{120}) due to (\ref{StarAdd}), and according to \cite[Section III.A]{C.Nair:2014}, with continuity, we have (\ref{121}).
\end{enumerate}
From (\ref{Theo1mieq1}), (\ref{LiuTie}), (\ref{Theo1ieq8}), (\ref{Theo1ieq9}) and (\ref{MIMO01}), we have
\begin{align}
%R_1+\mu R_2 &\le \lambda(C_1+C_2)+\frac{\mu-\lambda}{2}  \log \left(b^2 P_1+P_2+2 b \rho^* \sqrt{P_1 P_2}+1\right) \nonumber\\
%& \quad - \frac{\lambda}{2} \log \left(\frac{b^2 P_1+P_2+2 b \rho^* \sqrt{P_1 P_2}+1+N_3}{1+N_3} \right)  + \frac{\lambda}{2}\log\frac{(1-\rho^{*2})b^2P_1+1+N_3}{1+N_3} \nonumber\\
%& \quad+ \frac{\lambda}{2}\log\frac{(1-\rho^{*2})P_2+1+N_3}{1+N_3} +(1-\lambda)C_{\text{MIMO}12}^{\frac{\mu-\lambda}{1-\lambda}}(\rho^*) \nonumber\\
%& \quad-\frac{\mu-\lambda}{2} \log (b^2 P_1+P_2+1+2b\rho^* \sqrt{ P_1 P_2}) \nonumber\\
%&= \lambda(C_1+C_2) +(1-\lambda)C_{\text{MIMO}12}^{\frac{\mu-\lambda}{1-\lambda}}(\rho^*) \nonumber\\
%&\quad +\frac{\lambda}{2}  \log \frac{((1-\rho^{*2})b^2P_1+1+N_3)((1-\rho^{*2})P_2+1+N_3)}{(1+N_3)(b^2 P_1+P_2+2 b \rho^* \sqrt{P_1 P_2}+1+N_3)}. \label{Plug}
2(R_1+R_2) &\le (C_1+C_2)+\frac{1}{2}  \log \left(b^2 P_1+P_2+2 b \rho^* \sqrt{P_1 P_2}+1\right)\nonumber\\
& \quad - \frac{1}{2} \log \left(\frac{b^2 P_1+P_2+2 b \rho^* \sqrt{P_1 P_2}+1+N_3}{1+N_3} \right)  + \frac{1}{2}\log\frac{(1-\rho^{*2})b^2P_1+1+N_3}{1+N_3} \nonumber\\
& \quad+ \frac{1}{2}\log\frac{(1-\rho^{*2})P_2+1+N_3}{1+N_3} +C_{\text{MIMO}}^{\text{sum}}(\rho^*) \nonumber\\
& \quad-\frac{1}{2} \log (b^2 P_1+P_2+1+2b\rho^* \sqrt{ P_1 P_2}) \nonumber\\
&= (C_1+C_2) +C_{\text{MIMO}}^{\text{sum}}(\rho^*) \nonumber\\
&\quad +\frac{1}{2}  \log \frac{((1-\rho^{*2})b^2P_1+1+N_3)((1-\rho^{*2})P_2+1+N_3)}{(1+N_3)(b^2 P_1+P_2+2 b \rho^* \sqrt{P_1 P_2}+1+N_3)}. \label{Plug1}
\end{align}
The above is true for any $N_3  \geq 0$. Take $N_3$ as
\begin{align}
N_3=b \sqrt{P_1 P_2}\left(\frac{1}{\rho^*}-\rho^*\right)-1. \label{T1N3}
\end{align}
It can be seen that the value of $N_3$ in (\ref{T1N3}) is non-negative because we consider the case where $\rho^*$ satisfies (\ref{rhoCond}) if $b \geq 0$ and where $\rho^*$ satisfies (\ref{rhoCond02}) if $b <0$.
Plugging (\ref{T1N3}) into (\ref{Plug1}), 
we obtain
\begin{align}
2(R_1+R_2) &\le \left(C_1+C_2-\frac{1}{2} \log \frac{1}{1-\rho^{*2}} \right) +C_{\text{MIMO}}^{\text{sum}}(\rho^*) \nonumber\\
&= f_C(\rho^*) +C_{\text{MIMO}}^{\text{sum}}(\rho^*). \label{finished02}
%&= 2f_C^{12}(\rho^*)\label{finished02}
\end{align}
Due to symmetry, we may swap the indices 1 and 2 and re-derive the formulas from (\ref{rhoCond}) to (\ref{finished02}), and obtain that when $\rho^*$ satisfies 
\begin{align}
0 < \rho^* \leq \sqrt{1+\frac{1}{4a^2P_1P_2}}-\sqrt{\frac{1}{4a^2P_1P_2}} \nonumber
\end{align}
in the case of $a>0$, and if $\rho^*$ satisfies
\begin{align}
-\sqrt{1+\frac{1}{4a^2P_1P_2}}+\sqrt{\frac{1}{4a^2P_1P_2}} \leq  \rho^* <0 \nonumber
\end{align}
in the case of $a<0$, 
we again have (\ref{finished02}). 
%
%
%%%%%%%%%%%
%
Thus, from (\ref{theother}) and (\ref{finished02}), we have proved Theorem \ref{ub2}.

%%%%% APPENDIX B

\bibliographystyle{unsrt}
\bibliography{Thesis}

\end{document}